\documentstyle[epsf,aps,preprint]{revtex}
\begin{document}

\title{Like-Charge Attraction through Hydrodynamic Interaction}

\author{
Todd M. Squires$^1$ and Michael P. Brenner$^2$}

\address{
$^1$Department of Physics, Harvard University, Cambridge, MA 02138\\
$^2$Department of Mathematics, MIT, Cambridge, MA 02139}

\maketitle
\begin{abstract}
We demonstrate that the attractive interaction measured
between like-charged colloidal spheres near a wall can be
accounted for by a nonequilibrium hydrodynamic effect.  We present both analytical
results and Brownian dynamics simulations which quantitatively capture the 
one-wall experiments of
Larsen and Grier (Nature {\bf 385}, 230, 1997). 
\end{abstract}
\newpage
Colloidal spheres provide a simple model system for understanding the
interactions of charged objects in a salt solution. Hence, it came as a
great surprise when it was observed that two like-charged spheres can
attract each other when the spheres are confined by walls \cite{kep94,car96,croc96,lar97}. 
Since both the charge
densities and sizes of the spheres in question are in the range of large
proteins, it would be expected that a change in sign of this interaction
would have important implications for biological
systems \cite{honig}.  Theorems by Sader
and Chan\cite{sad99} and Neu\cite{neu99} demonstrate that under very general
conditions the Poisson-Boltzmann equation for the potential between
like-charged spheres in a salt solution will not admit attractive interactions. 
Explanations
for the observed attraction have thus exclusively focused on deviations from
the classical Derjaguin, Landau, Verwey and Overbeek (DLVO) theory.

Herein, we propose that an attractive interaction of two like-charged
colloidal spheres measured in the presence of a single wall can arise from a
non-equilibrium hydrodynamic effect. The idea is that the relative motion 
between two
spheres depends on {\sl both} the forces acting between them and in
addition, their hydrodynamic coupling. 
In a bulk solution, far from solid boundaries, an external force
acting on two identical spheres cannot change their relative positions.
This is a consequence of the kinematic reversibility of Stokes flow and of the 
symmetries inherent in the problem.

However, these symmetries are broken in confined geometries, where the
hydrodynamic effect of boundaries is important. In this situation, relative
motion between the particles could stem from {\sl either} an interparticle
force, {\sl or} from a hydrodynamic coupling caused by forces acting on each
of the particles individually. In a typical experiment with charged
colloidal spheres, the charge density on the walls of the cell is of order
the charge density on the spheres\cite{grier}. We demonstrate that the
hydrodynamic coupling between two spheres caused by their repulsion from a
wall leads to motion which, if interpreted as an equilibrium property, is
consistent with an effective potential between the spheres with an
attractive well. Our calculations quantitatively reproduce the experimental
measurements of these potentials.

The response of a particle to an external force is significantly changed near
a wall because the flow field must vanish identically on the wall. For point 
forces, Lorentz determined this wall-corrected flow field \cite{lor}, which 
Blake later expressed using the method 
of image forces \cite{blake}, analogous to image charges used in
electrostatics.  Images of the appropriate strength on the opposite side of
the wall exactly cancel out the fluid flow on the wall. When two particles
are pushed away from a wall, the flow field from one particle's image tends
to pull the other particle towards it, and vice versa (Fig. 1). This
decreases the distance between the particles. 
\begin{figure}[tbp]
\centerline{\epsfysize=3.5in\epsfbox{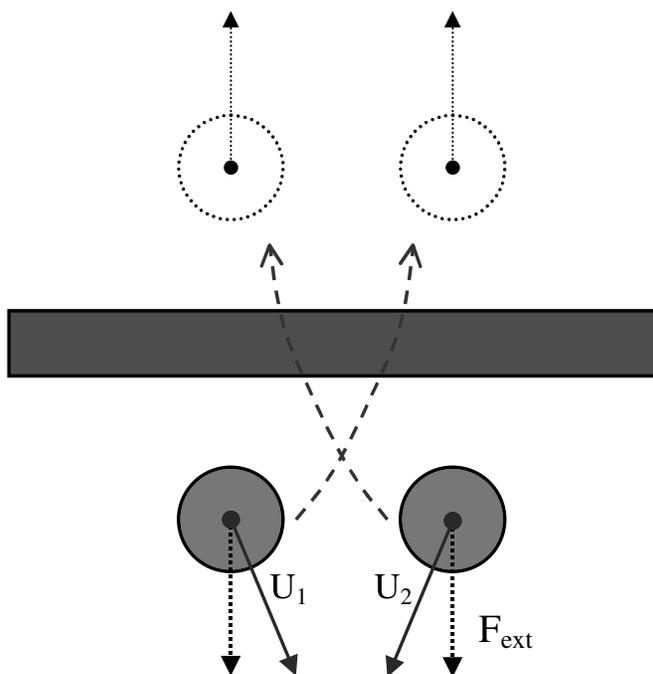}}
\caption{ Two spheres forced away from a wall are drawn together by
hydrodynamic coupling, because the image force of one particle pulls the
other particle towards it.}
\end{figure}

The attractive interaction between two charged spheres in the presence of a
wall can now be understood with a simple picture. When the spheres are 
sufficiently 
close to the wall, they are electrostatically repelled from it. The net force on
each sphere thus includes both their mutual electrostatic repulsion and
their repulsion from the wall. How the spheres respond depends on their 
hydrodynamic
mobility:  when the spheres are close together (Fig. 2a), their
mutual repulsion overwhelms any hydrodynamic coupling, and the spheres will
separate as expected for like-charged bodies. However, when they are beyond 
some critical separation
(Fig. 2b), the hydrodynamic coupling due to the wall force overcomes
the electrostatic repulsion, so that the particles move together as they
move away from the wall. 

\begin{figure}[tbp]
\centerline{\epsfysize=2.0in\epsfbox{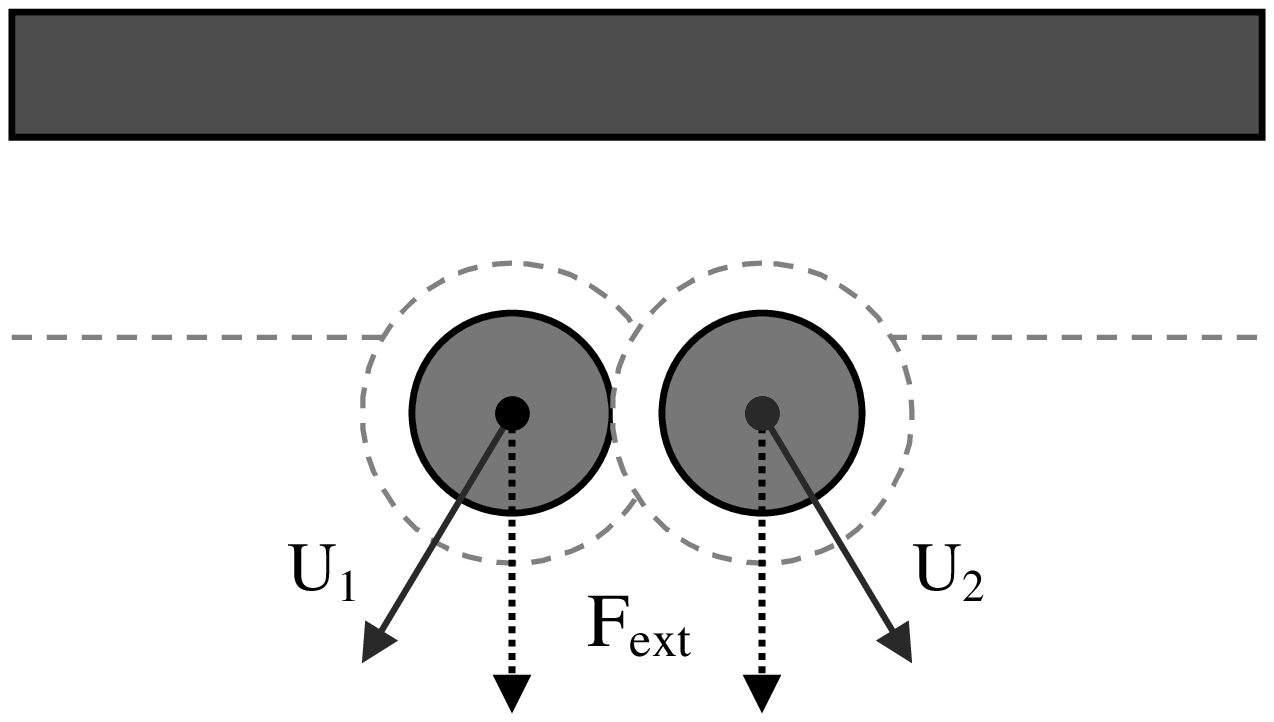}
\epsfysize=2.0in\epsfbox{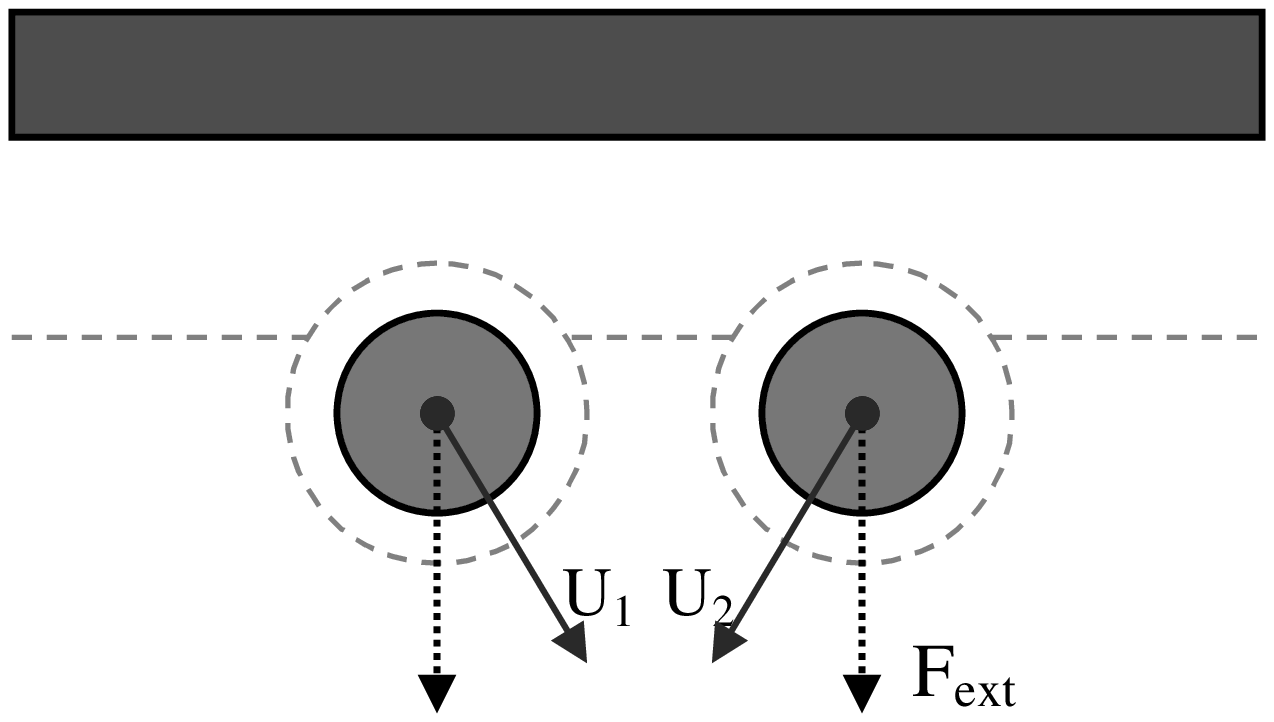}}
\caption{(A) If the screening clouds of the two spheres overlap
sufficiently, the mutual repulsion is stronger than the hydrodynamic
coupling. (B) When the spheres are further apart, the hydrodynamic coupling
dominates.}
\end{figure}

Although this decrease in mutual separation is a non-equilibrium kinetic effect, 
it could be interpreted 
as the result of an attractive equilibrium pair-potential. This is most clearly 
understood without Brownian motion. Two particles
initially located a distance $r$ apart move because of both interparticle forces
and the repulsive force from the wall. The response of these two particles
to forces ${\bf F}_{1}$ and ${\bf F}_{2}$ is expressed by the hydrodynamic mobility 
tensor $ {\bf b}\left( {\bf X}_{1},{\bf X}_{2}\right) $, defined by 
\begin{equation}
{\bf v}={\bf b}\left( {\bf X}_{1},{\bf X}_{2}\right) \cdot {\bf F},
\end{equation}
where ${\bf v}=({\bf \dot{X}}_{1},{\bf \dot{X}}_{2})$ are the particle
velocities and ${\bf F}=({\bf F}_{1},{\bf F}_{2})$ are the forces on the particles. 
Thus, the distance between the spheres (measured in the plane parallel to the walls)
will change by an amount $\Delta r = \Delta x_2 - \Delta x_1$ in a small time
 $\Delta t$, where we denote the $x$-direction to be along the line connecting the
spheres, and the $z$-direction to be perpendicular to the wall. 
Utilizing symmetries of the mobility tensor, it is straightforward to show that
$\Delta r$ will be 
\begin{equation}
\Delta r=\left\{ 2 (b_{X_2 X_2}-b_{X_2 X_1}) |F_p| + 2 b_{X_2 Z_1} F_w
\right\} \Delta t,
\end{equation}
where $F_p$ and $F_w$ are respectively the repulsive electrostatic sphere-sphere and 
sphere-wall forces.  The tensor component $b_{X_2Z_1}$ refers 
to the $x$-motion of particle 2
due to a force in the $z$-direction on particle 1, and so on.  

If this system were assumed to be in equilibrium, then the relative motion would be
interpreted as the result of an effective potential, so that an effective force 
$F_{eff}=-\partial_rU_{eff}$ 
\begin{equation}
\Delta r=\left\{ 2 (b_{X_2X_2}-b_{X_2 X_1}) |F_{eff}| \right\} \Delta t,
\end{equation}
so that one would determine this effective potential to be given by 
\begin{equation}
U_{eff}(r,h)=U_{p}(r)-
F_w\int_{\infty }^{r}\frac{b_{X_2 Z_1}(r,h)}{b_{X_2 X_2}(h)-b_{X_2 X_1}(r,h)}dr,  \label{analytic}
\end{equation}
where $U_p(r)$ is the interparticle thermodynamic pair potential, $r$ is the separation between 
particles, and $h$ is their distance from the wall.

In order to compare our results with experiments, we determine the
hydrodynamic mobilities in the point-force limit, using Blake's solution
\cite{blake}.  We use the DLVO potential 
\cite{saville,der,ver48} for the electrostatic interaction of two spheres  
in the form presented by Larsen and Grier \cite{lar97}, 
\begin{equation}
\frac{U_{DLVO}}{k_{B}T}=Z^{2}\lambda _{B}\left( \frac{e^{\kappa a}}{1+\kappa
a}\right) ^{2}\frac{e^{-\kappa r}}{r},
\label{dlvogrier}
\end{equation}
where $a$ and $Z$ are respectively the radius and effective charge of each sphere, the
Bjerrum length $\lambda _{B}=e^{2}/\varepsilon k_{B}T$, and the
Debye-H\"{u}ckel screening length $\kappa ^{-1}=(4\pi n\lambda _{B})^{-1/2},$
with a concentration $n$ of simple ions in the solution. This formula is 
obtained 
using effective point charges in a linear superposition approximation.  
To determine the repulsive electrostatic force between each sphere and the wall, we 
used the same effective point-charge approach to obtain
\begin{equation}
\frac{U_{wall}}{k_{B}T}=Z\sigma _g\lambda _{B}\frac{e^{\kappa a}}{\kappa
(1+\kappa a)}e ^{-\kappa h},
\label{wallforce}
\end{equation}
where $\sigma_g$ is the effective charge density on the glass wall.
We note that while the functional form of this equation is correct, it is not 
clear that the effective charges in equations
(\ref{dlvogrier}) and (\ref{wallforce}) will be exactly the same, 
as geometric factors buried in each effective charge will vary from situation to 
situation.  A more reliable description of sphere-sphere and wall-sphere 
interactions will be necessary for quantitative comparisons with
independently measured charge densities.

Using all of Larsen and Grier's experimental parameters as inputs to the theory, we 
numerically integrate (\ref{analytic}) to obtain this apparent
effective potential.  The only necessary parameter not given is the surface charge
density of the glass walls $\sigma_g$, which we take to be $\sigma_g = 5 \sigma_p$,
consistent with Kepler and Fraden's measurements \cite{kep94}. Fig. 3 shows this 
effective potential for various sphere-wall separations.
The hydrodynamic coupling of collective motion away from the wall with
relative motion in the plane of the wall leads to an attractive component.
It is important to emphasize that this hydrodynamic coupling is a {\it %
kinematic} effect, and has no thermodynamic significance--all forces acting
on the spheres are purely repulsive. 
\begin{figure}[tbp]
\centerline{\epsfysize=3.5in\epsfbox{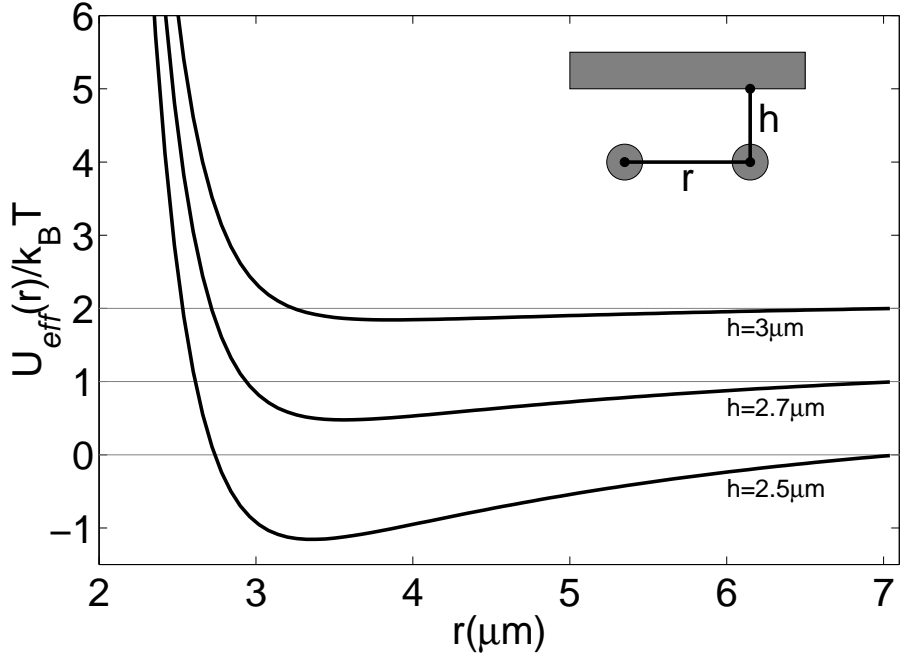}}
\caption[]{Plot of the analytic effective potential (\ref{analytic}) for three 
different wall separation distances. The simulated effective potentials 
(Fig. 4) are slightly 
shallower because the pair of spheres drifts off the wall into areas with a 
shallower
 well.}
\end{figure}

We note as well that a simple approximate expression exists for the
hydrodynamic term in the effective potential (\ref{analytic}),
since $b_{X_2 X_2}(h)/b_{X_2 X_1}(r,h)\sim O(h/a) >>1.$
Approximating the denominator in the integrand as simply $b_{X_2 X_2}$, 
we explicitly evaluate the integral to give
\begin{equation}
U_{eff}(r,h)=U_{p}(r)-\frac{F_w}{1-\frac{9 a}{16 h}}\frac{3 h^3 a}{(4h^2 +
r^2)^{3/2}}.
\end{equation} 

As a complement to this analytic approach, we simulate the dynamics of this system, 
using (\ref{dlvogrier}) and (\ref{wallforce}) for the sphere-sphere and wall-sphere
forces, respectively.  We account for Brownian motion of the particles in the standard
Stokes-Einstein fashion, whereby the diffusion tensor is proportional to the
mobility tensor, ${\bf D}=k_{B}T{\bf b}$ \cite{bat76,erm78,hinch}.  Using
all experimental parameters and $\sigma_g = 5 \sigma_p$ as explained above, 
we performed a computer version of Larsen and Grier's
experiment, and analyzed the resulting data using their methods\cite{croc96b}.
Our results suggest that this approach includes all of the essential
ingredients necessary for quantitatively understanding their observations.

In Fig. 4, we present simulations for the two cases presented by Larsen and Grier: the
first with the spheres $2.5$ microns from the wall, so that they interact
significantly with the charge double layer of the wall, and the second
starting $9.5$ microns from the wall, well outside of the wall's charge
double layer.

\begin{figure}[tbp]
\centerline{\epsfysize=2.5in\epsfbox{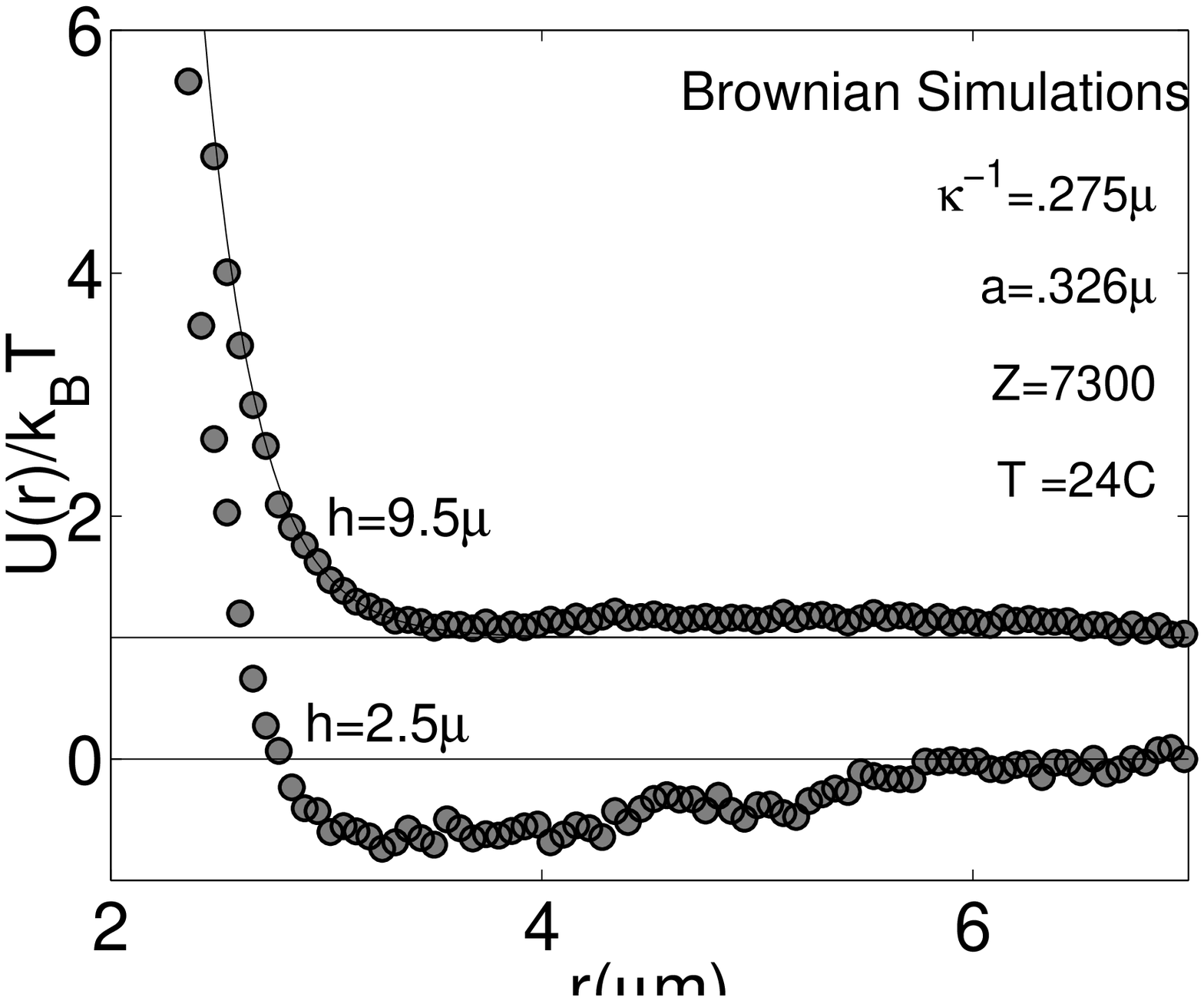}
\epsfysize=2.5in\epsfbox{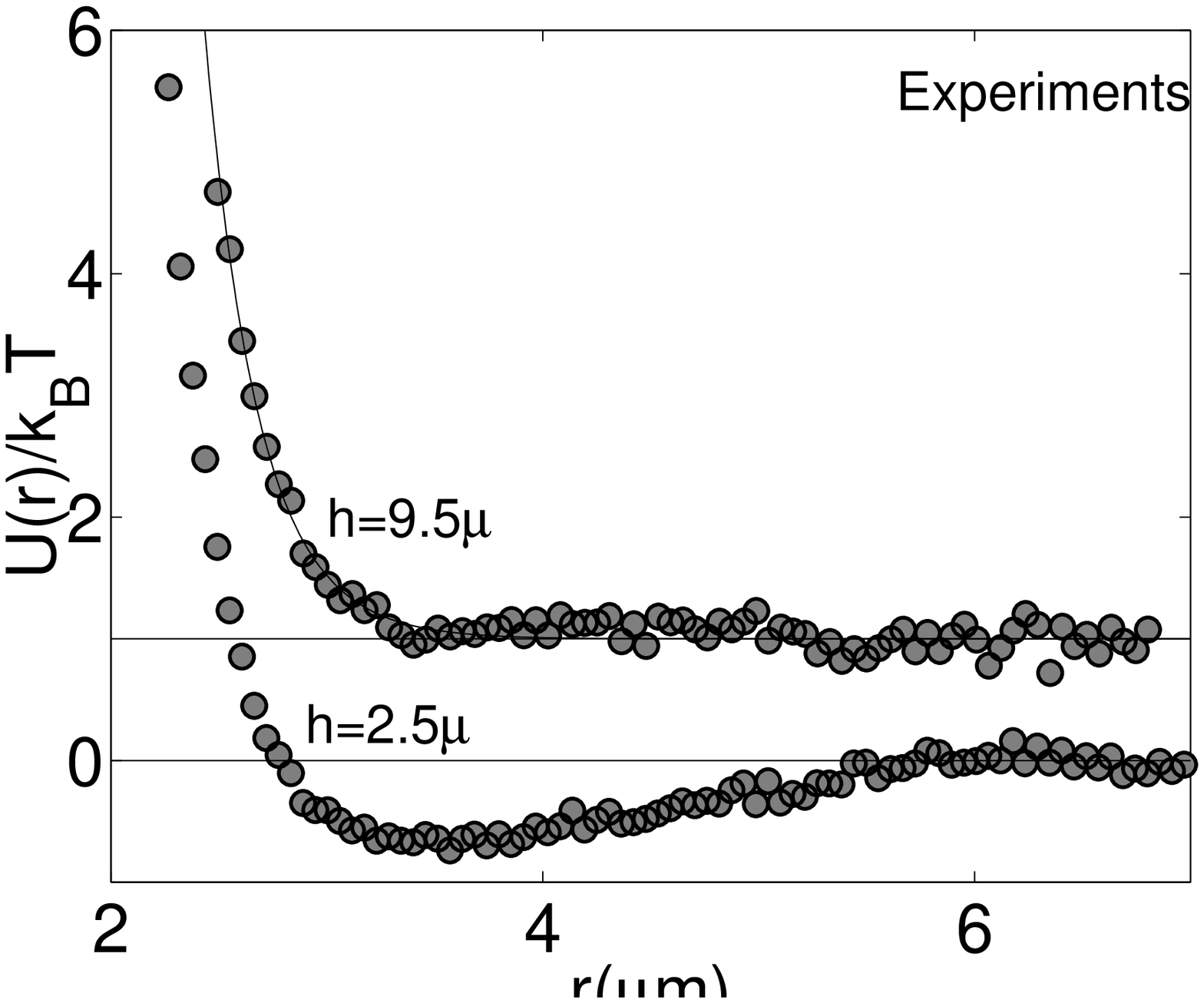}}
\caption[]{Comparison between Brownian dynamics simulations and experiments%
\protect\cite{lar97} for the effective potential between two colloidal
charged spheres near a wall. Two situations are presented: spheres close to 
the wall (h=2.5 $\mu$), and far from the wall (h=9.5 $\mu$).  These are offset by
1 $k_B T$ for clarity.  The simulations were carried out using standard
methods \protect\cite{erm78,hinch}, taking all parameters for the DLVO
potential as those measured in the experiments \protect\cite{lar97}. The
simulations were analyzed using the same techniques used in the experiments 
\protect\cite{croc96b}. The only parameter that is not precisely measured is
the charge density on the wall, which we take to be $\sigma_g=5\sigma_p$.}
\end{figure}

Our theoretical picture agrees quantitatively with measured data. Moreover,
there are many consequences of the theory that can be tested experimentally:
(1) Effective kinetic potentials can be predicted for different sets of
conditions and quantitatively compared with experiments; (2) The hydrodynamic
mechanism requires a net drift of the particles away from the wall, which could 
be independently measured. 
(3) Finally, the theory
provides a simple explanation for the observation that the attraction
disappears when the salt concentration is increased. While this at
first seems counterintuitive--the particles are mutually attractive only
when they are mutually repulsive--the significance of the wall-driven
hydrodynamic coupling makes this clear. 

Several pieces of experimental evidence have been collected which seemed to 
suggest
the existence of an attractive minimum in the thermodynamic pair potential
of like-charged colloidal particles in confined geometries. Besides the one
wall experiment under discussion, attractive pair potentials have been
observed for two spheres trapped between two walls\cite{croc96}, and for a
suspension of spheres trapped between two walls\cite{kep94,car96}. In
addition, it has been shown that metastable colloidal crystals take orders
of magnitude longer to melt than would be expected without a thermodynamic
attraction\cite{largrier96}.  Similarly, voids in colloidal crystals take much 
longer to close than expected\cite{ise}.  It is not clear how the theory presented
here will bear upon these experiments.

The theory presented in this paper offers a non-equilibrium hydrodynamic
explanation for the attractive potential in the single-wall experiments
without invoking a novel thermodynamic attraction. We have found
quantitative agreement with experimental results when the effective wall
charge density is chosen to be $\sigma_g = 5\sigma_p$, which is in the
ballpark of measured estimates.  Without a quantitative measurement of this
parameter, this work does not strictly rule out the
possibility that a novel attraction exists. This situation can be
definitively resolved by more quantitative comparisons with experiments.

{\bf Acknowledgments:} We are indebted to D. Grier and E. Dufresne for
introducing us to their experiments, and for a stimulating collaboration.
Useful discussions with J. Crocker, H. Stone, and D. Weitz are gratefully
acknowledged. This research was supported by the Mathematical Sciences
Division of the National Science Foundation, the A.P. Sloan Foundation, and
the NDSEG Fellowship Program (TS).

\end{document}